\documentclass[final,1p,times]{elsarticle} % do not change
\usepackage{graphicx} % do not change
\usepackage{amssymb} % do not change
\usepackage{amsthm} % do not change
\usepackage{lineno} % do not change

\usepackage{amsmath} 

\newcommand{\mytext}[1]{\text{ #1}}

\newcommand{\GeV}{\ensuremath{\mytext{GeV}}}
\newcommand{\GeVc}{\ensuremath{\mytext{GeV}/c}}
\newcommand{\pT}{\ensuremath{p_{T}}}
\newcommand{\MeanpT}{\ensuremath{\langle p_{T}\rangle}}
\newcommand{\kT}{\ensuremath{k_{T}}}
\newcommand{\ET}{\ensuremath{E_{T}}}

\journal{Nuclear Physics A} % do not change
\begin{document} % do not change

\begin{frontmatter} % do not change

%% QM09Author: please enter your  
%% Title, author and address info here; please do not use footnotes

% Your Title - please insert
\title{Jet Studies in STAR via Di-jet Triggered (2+1) Multi-hadron Correlations}

% Principle author, and co-authors - please insert
\author{Kolja Kauder for the STAR Collaboration}

% Address - please insert
\address{University of Illinois at Chicago, Chicago, IL 60607, USA}

\begin{abstract} % do not change
%% Text of abstract goes here - please insert
  We explore jet-medium interactions via the recently developed 
  multi-hadron correlation or ``2+1'' technique. We restrict the di-jet kinematics by 
  selecting a pair of approximately back-to-back high $\pT$ hadron triggers and study associated particles.
 Here we present our study of di-jet systematics comparing
  the measurements of associate yields and spectra
  in 200\GeV Au+Au and d+Au collisions in two different scenarios.
  We present $\sum \pT$ as an estimate for energy loss. First findings indicate little to no 
  energy loss in the symmetric ``2+1''  scenario whereas model predictions are about 2\GeV.

\end{abstract} % do not change

\end{frontmatter} % do not change

%% QM09: we keep linenumbers at least for initial version
%% \linenumbers % do not change

%% start of main text - please insert. 
\section{Introduction}
\label{sec:introduction}
A strongly interacting partonic medium
is evident in multiple observables at RHIC.
One manifestation of this medium, the
 jet-quenching effect, was found in single-particle
spectra \cite{PHENIX:RAA} as well as in away-side di-hadron angular correlations \cite{adams:152301}.
The latter have already yielded new insights into the
medium properties: away-side quenching indicates that the medium is
dense, and transverse momentum distributions of associated hadrons
are consistent with thermalization \cite{adams:152301}.
Alteration of the angular distribution on both 
(same- and away-) sides of a high \pT\,trigger reveals 
a strong modification of known hadron production mechanisms.
Alternatively, it has been
argued that the peculiar features found in the triggered
di-hadron distributions, such as a double-humped away-side structure \cite{Horner:2007gt}
and a ``ridge'' along pseudo-rapidity $\Delta\eta$ formed on the same-side of the trigger,
could be caused by the medium's response to propagated jets \cite{Putschke:2007mi}.

Several three-particle correlation techniques 
have been developed to further investigate proposed mechanisms for jet modifications like 
jet deflection or the Mach cone/shock wave mechanism.
Examples include $\Delta\phi$--$\Delta\phi$ correlation analyses \cite{Ulery:2006ha}
 and $\Delta\eta$--$\Delta\eta$ studies \cite{STAR:EtaEta}.
Both methods use one high \pT\,trigger particle and then study the correlation of two lower \pT\,associate particles, and in that sense 
are denoted as 1+2 techniques. With our 2+1 method we instead use a pair of correlated high \pT\,hadron triggers
to determine the jet axis and study associated hadron distributions with respect to this axis.
Thus we  aim to explore both same- and away-side features with a well-defined jet axis at the same time.
This new approach has already shown unique results \cite{Barannikova:2008zz}; here we present the most recent developments.

\section{Data Analysis}
\label{sec:data-analysis}
A description of the method and initial results were introduced in \cite{Barannikova:2008zz}. 
To summarize, the re-emergence of an away-side peak in angular correlations of two high \pT\,particles \cite{Adams:2006yt}
allows us to determine the away-side jet direction in both azimuth and pseudo-rapidity.
The highest transverse momentum particle in the event is designated as a primary trigger (T1)
if it is at mid-rapidity, $|\eta|<1$,  and if it satisfies the transverse momentum threshold.
A second trigger (T2) is selected back-to-back with the primary trigger, i.e.\,within $|\Delta\phi-\pi|<\alpha$,
where $\Delta\phi$ is the relative azimuthal angle between the two triggers.
We choose $\alpha = 0.2$ as a characteristic away-side peak width for this kinematic regime.
T2 satisfies the same pseudo-rapidity cuts
and both triggers have to obey additional track quality cuts.
The distribution of associated hadrons (A) is then studied in both $\Delta\eta$ and $\Delta\phi$ dimensions
with respect to the jet axis as determined by this pair of triggers. The associated hadrons have to satisfy similar track quality cuts as the trigger pair.

The raw angular correlation is constructed similarly to other di-hadron analyses \cite{Horner:2007gt,Putschke:2007mi}.
To correct for pair acceptance we use the mixed-event technique.
The combinatorial background is modulated by the $\cos(2\Delta\phi)$ function to account for elliptic flow effects.
We use published STAR results for the centrality and \pT \,dependence of the elliptic flow coefficient $v_2$ \cite{Adams:2004bi}.
The overall background level is determined via the Zero-Yield at Minimum (ZYAM) method as in \cite{Ulery:2006iw}.

Specific to 2+1 correlations is the correlated background, resulting from randomly associated trigger pairs.
The angular distribution of the associated hadrons for these trigger pairs is obtained from
 di-hadron correlations with appropriate kinematic selections.
We scale this background by the signal-to-noise ratio in the away-side peak trigger-trigger
 correlation used for jet axis determination and subtract this contribution from the raw distribution.
The resulting corrected signal is scaled back so that ultimately the correlation is normalized per di-jet trigger.

\section{Results}
\label{sec:results}

\begin{figure}[h!]
  \centering
  \includegraphics[scale=0.29]{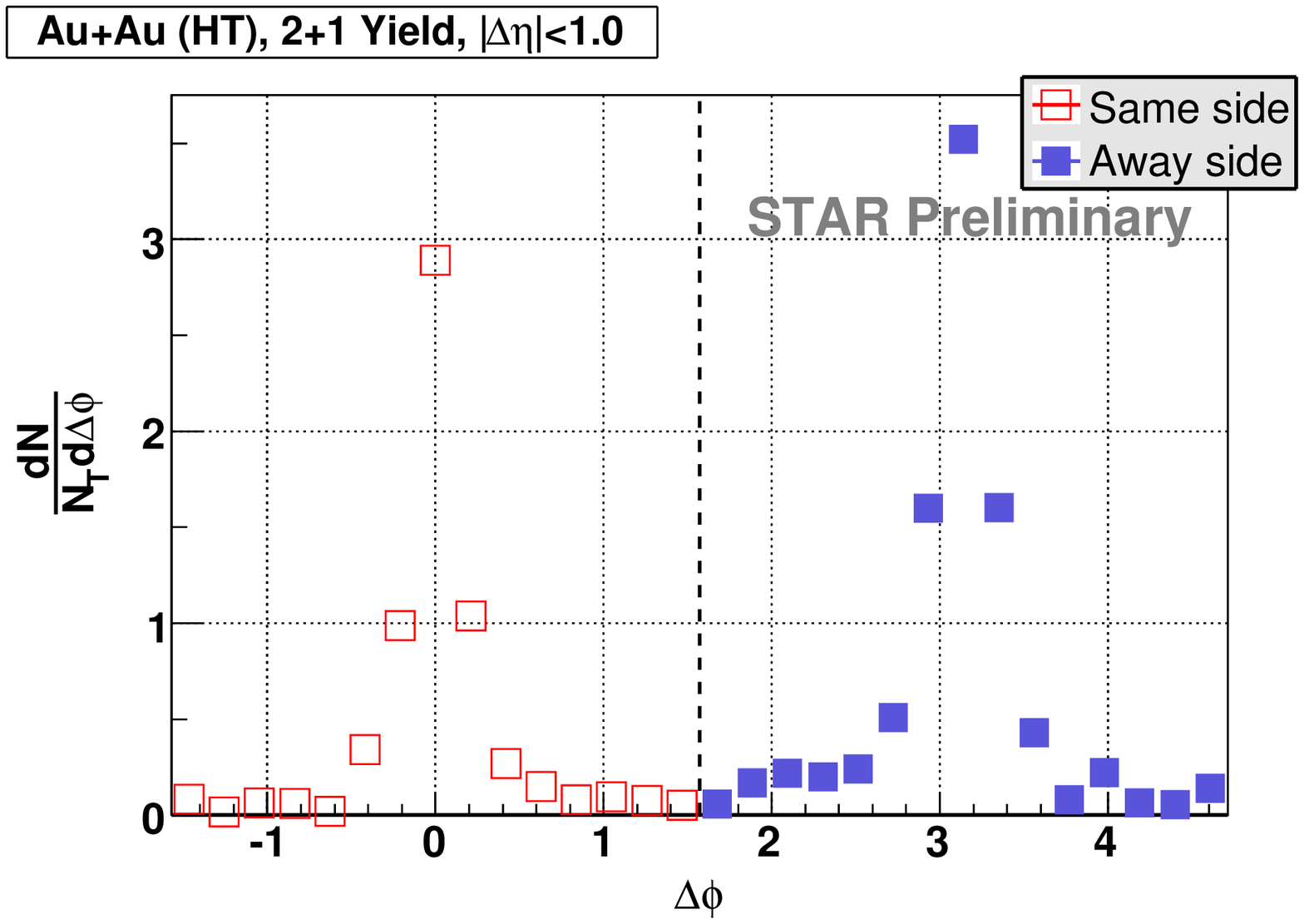}
  \includegraphics[scale=0.29]{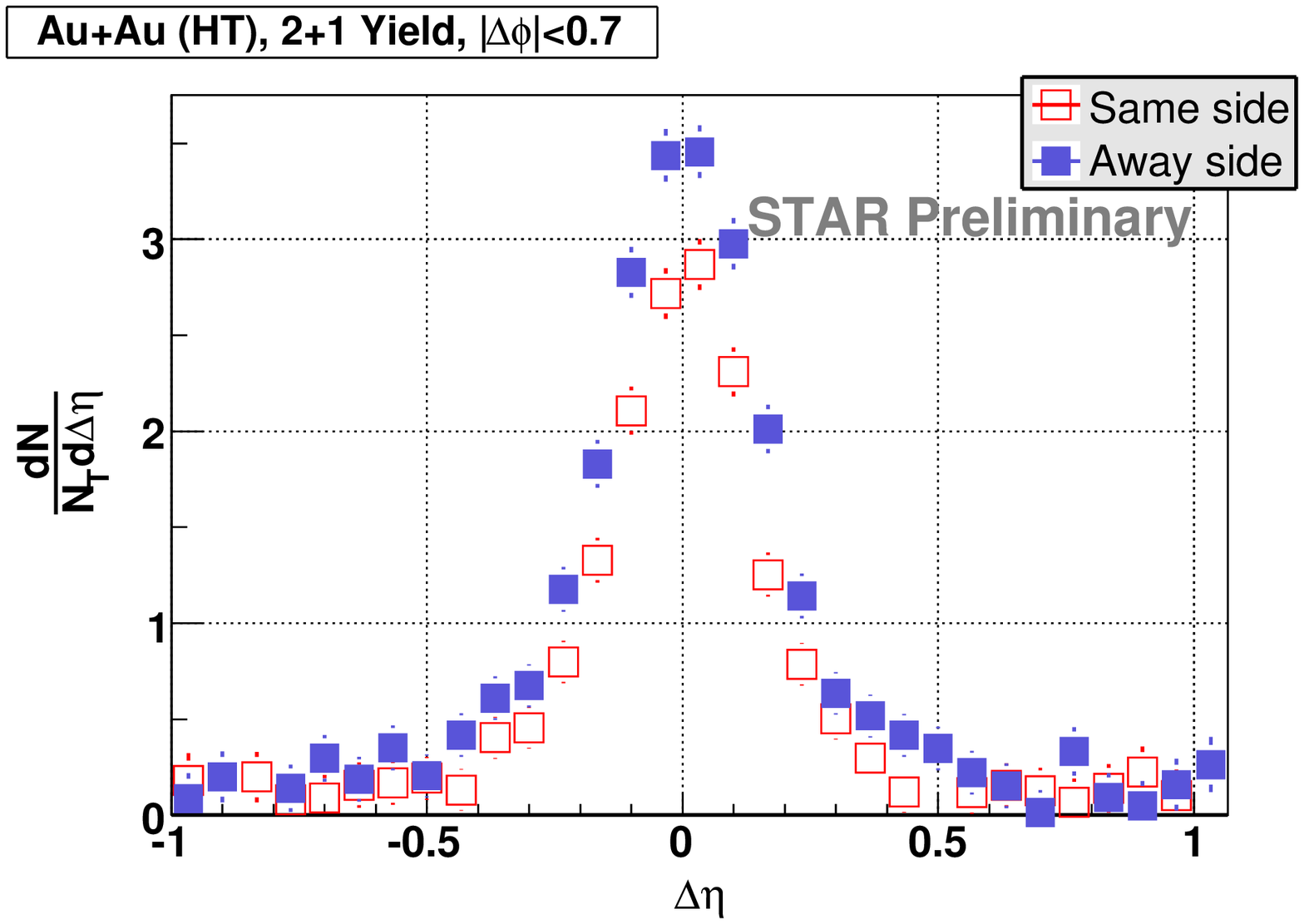}
\caption{Yields in the asymmetric case (normalized per number of triggers N$_T$): High-Tower triggered 200\GeV\,Au+Au data,
    $8\,\GeV<\ET (T1)<15\,\GeV$, $4\,\GeVc<\pT (T2)<\min\{\pT (T1),10\,\GeVc\}$, $1.5 \, \GeVc <\pT (A)<10\,\GeVc$.
    Upward triangles signify the same-side correlation and are compared directly with
    the away-side correlation in the downward triangles.
    $\Delta\phi$ and $\Delta\eta$ are calculated with respect to T1 on the same-side and with respect to T2 on the away-side.
    The left figure shows the $\Delta\phi$ projection; the right figure shows the $\Delta\eta$ projection.
    Error bars are statistical only.}
 \label{fig:AsymmRes}
\end{figure}

Below we consider  two scenarios with different relative \pT\,between the
two trigger particles. The first is a symmetric scenario, as in our early results \cite{Barannikova:2008zz}, in which
the T1 trigger is selected between 5 and 10 \GeVc\,
and T2 above 4 \GeVc. Associated hadrons were selected with \pT\,between 1.5 and 4 \GeVc.
20M 200\GeV\,Central Au+Au events were used for this analysis.
The availability of the High-Tower triggered 200 \GeV\,Au+Au data set in STAR with
 good statistics in the higher \pT\,region
enabled us to consider a second, asymmetric scenario. Using the Barrel Electromagnetic Calorimeter, we required the
T1 trigger's transverse energy \ET\,to be between 8 and 15 \GeV, rejecting triggers that coincided with a charged hadron track.
The selection criteria for T2 remained unchanged, associated particles were selected 
to have transverse momenta in the TPC above 1.5 \GeVc. 
Where \pT\,cuts were used, values above the resolution threshold of 10\GeVc\,disqualified an event in both scenarios.

Previously, we found no modifications on the away-side of a di-jet in the symmetric case \cite{Barannikova:2008zz}.
In the present analysis we confirmed the absence of the jet-medium modification effects for such di-jets
with more detailed studies of transverse momentum spectra of associated hadrons.
A tangential jet production scenario could account for such results.

In contrast, an asymmetric trigger pair should enhance long path lengths through the medium on the away-side.
The correlation structures resulting from a 2+1 analysis for the asymmetric triggers feature clear jet-like peaks
 on both same- and away-sides.
$\Delta\phi$ and $\Delta\eta $ projections of the 2-dimensional correlation function are shown in Figure \ref{fig:AsymmRes},
with no obvious broadening or indication of a ridge-like phenomenon.
In this new analysis we observe enhancement rather than suppression of away-side yield
in contrast to all di-hadron correlation results for this kinematic range of associated particles.
We next investigate the correlation peaks in more detail.

\begin{figure}[h!]
  \centering
  \includegraphics[scale=.29]{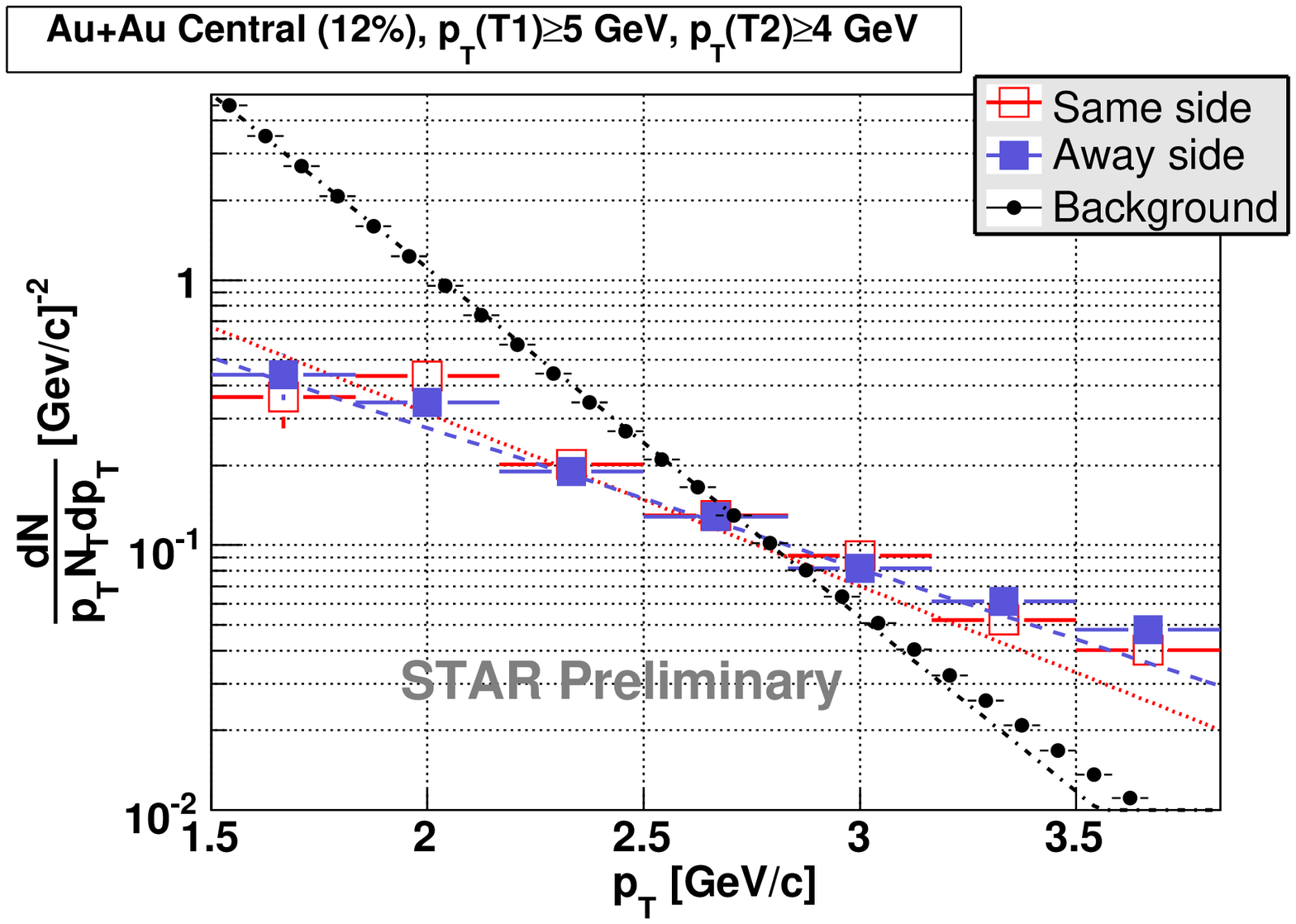}
  \includegraphics[scale=.29]{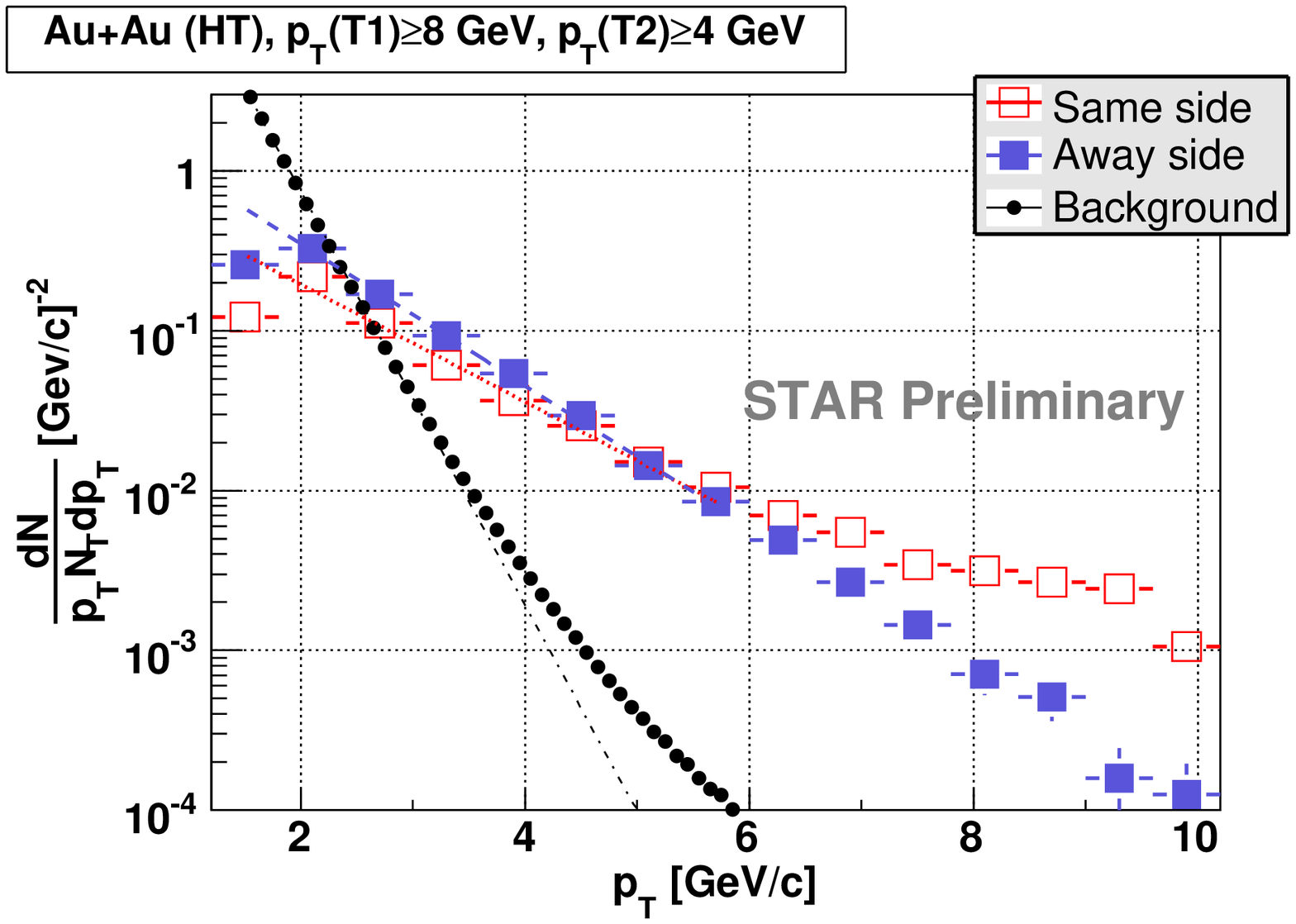}
\caption{
    $\pT$ spectra  (normalized per number of triggers N$_T$) for the symmetric case on the left and the asymmetric case on the right.
    Upward triangles signify the same-side correlation while 
    the away-side correlation is represented by downward triangles. In both cases the minimum bias inclusive 
    background is given as a reference. As a visual aid fits to $e^x/x$ are shown with dashed lines. Error bars are statistical only.
 }
  \label{fig:Spectra}
\end{figure}

Figure \ref{fig:Spectra} shows the \pT\,spectra of associated hadrons in the jet peaks as defined by $\left|\Delta\eta\right|<0.5$ and
$\left|\Delta\phi\right|<0.5$ for both cases.
Note the expected significant hardening of the associated hadron spectra on both sides of the primary trigger as
 compared to the inclusive hadron distribution.
Also, just as in the case of angular variables,
the spectra on both sides  are virtually identical for the symmetric case. 
On the other hand,
the away-side spectrum of the asymmetric trigger pair is softer than that of the same-side.
Comparison of symmetric di-jet spectra between Central Au+Au and minimum bias d+Au data (available from Run III)
confirmed the absence of modifications within our uncertainties.
Analysis of the d+Au data from Run VIII is in progress.

\begin{figure}[h!]
  \centering
  \includegraphics[scale=.29]{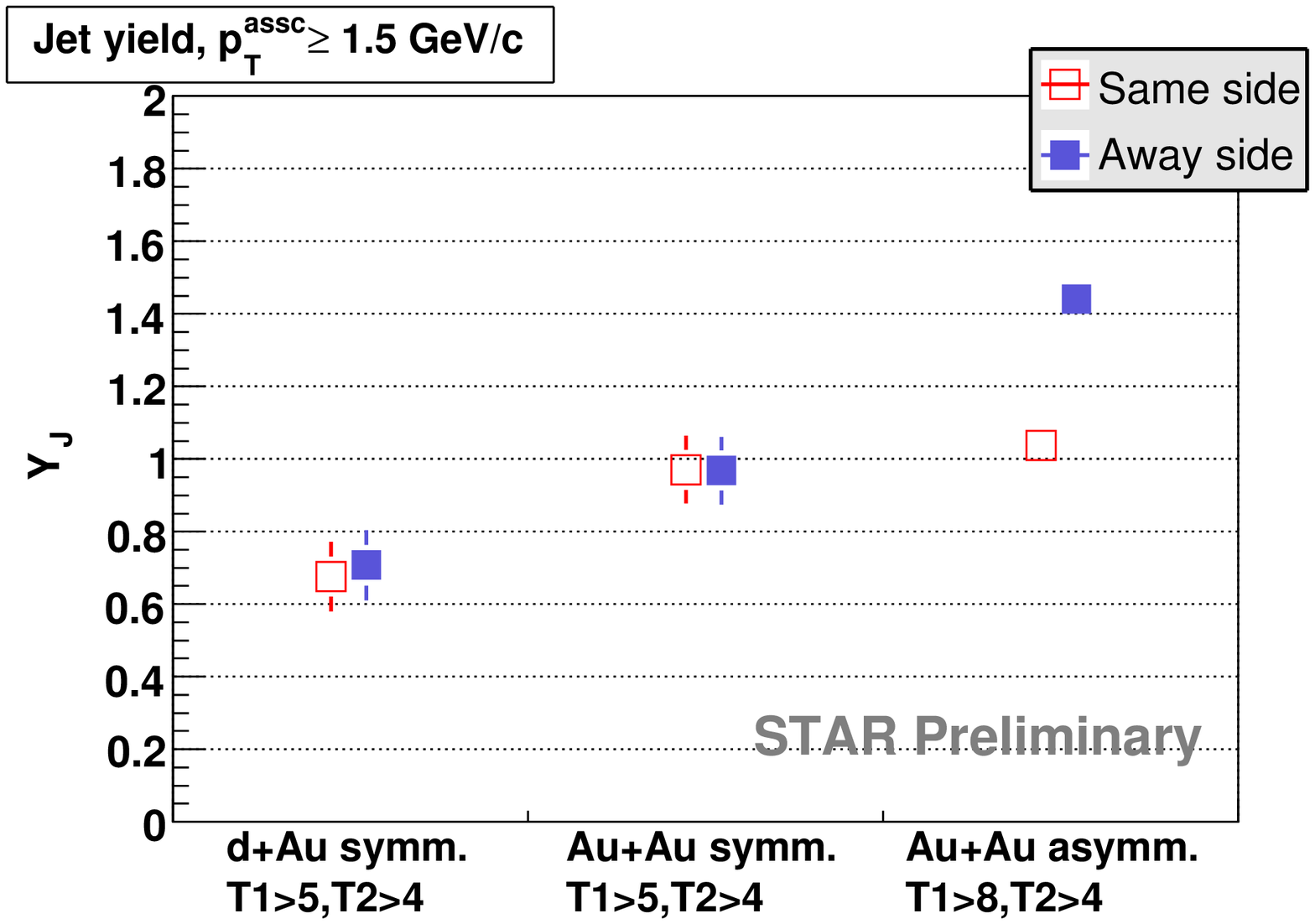}
  \includegraphics[scale=.29]{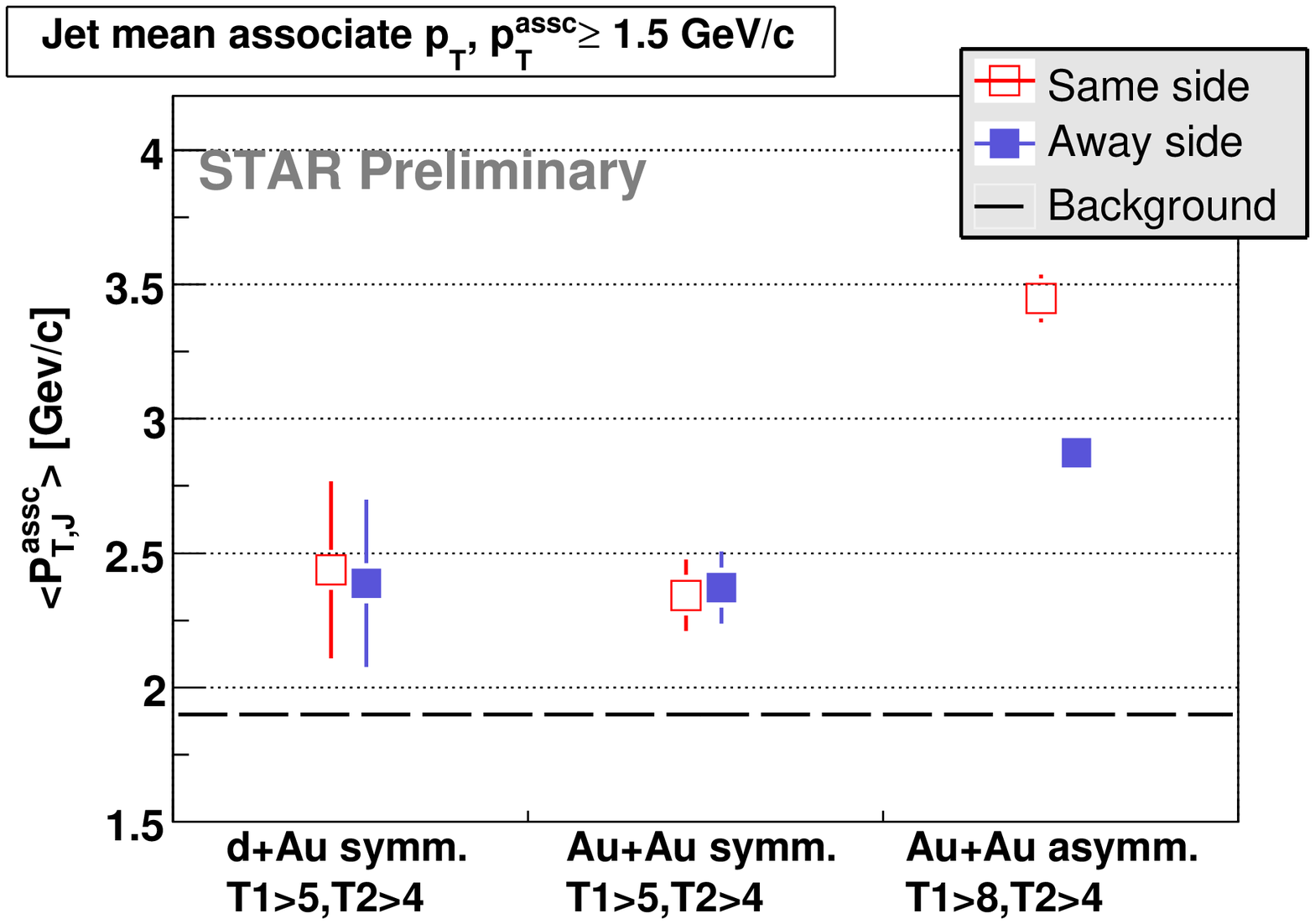}
  \caption{ Yield (left) and \MeanpT (right) in the symmetric and asymmetric case.
    For the symmetric case, the d+Au values are also shown.
    Upward triangles are used for the same-side correlation, 
    downward triangles for the away-side correlation.
   Error bars are statistical only.
  }
  \label{fig:Numbers}
\end{figure}

% To further quantify our findings the average associate \pT\,and the total yield in the jet peaks for both scenarios  
% are shown in Figure \ref{fig:Numbers}. 
To further quantify our findings the average \pT\,of associated hadrons and the total yield in the jet peaks for both scenarios  
are shown in Figure \ref{fig:Numbers}. 
We are finalizing the systematic uncertainties, but they are expected to be between 20\% and 30\% and
strongly correlated between same- and away-side. If this expectation holds, there will be no appreciable difference 
between d+Au and Au+Au systems in the symmetric case. 
By contrast, in the asymmetric case, we see
 a spread between same- and away-side. The enhancement in total yield is offset by an attenuation in 
\MeanpT.

In \cite{renk:014903} two models were used to compute 
% total energy deposition in the medium for di-hadron triggered events. As a proxy for this value we compute
% the scalar sum of trigger and associate \pT\,on both same- and away-sides and consider the difference between the two.
total energy deposition in the medium for di-hadron triggered events. As a proxy for this value we sum up
the \pT\,of trigger and associated particles on both same- and away-sides and consider the difference between the two.
In the symmetric case we find similar values of $1.50 \pm 0.31 \GeVc$ (Au+Au) and $1.64 \pm 0.35 \GeVc$ (d+Au),
consistent with
the estimate for \kT\,smearing used in \cite{renk:014903}.
The differences between the correlation peaks, if any, are not accessible
with these measurements.
For the asymmetric case we do not yet have a comparison with d+Au. A reasonable assumption is that the
\kT\,effect has about the same trigger-independent value of 1.6 \GeVc. Subtracting this from the
 Au+Au $\sum \pT$ difference of 4.3 \GeVc\,results in an estimated energy loss of 3\GeV\,in the fiducial range.

\section{Summary}
\label{sec:summary}

In correlations using  symmetric \pT\,trigger pairs we found no evidence for  modifications in yield
or $\pT$ spectra for both the 
Central Au+Au and d+Au collisions, and for same- and away-sides.
This is consistent with minimal medium interaction, e.g. 
strong surface bias in the Au+Au correlation. 
We expanded our 2+1 correlation studies
of di-jet properties to include an asymmetric trigger selection scenario
and again found the peaks on the same-side and away-side
to be similar in shape.
However, the away-side exhibited an increased yield 
balanced by a softer spectrum resulting in a lower \MeanpT.

We presented $\sum \pT$ as an estimate for energy loss, and our first results suggest negligible 
energy loss in the symmetric region
where model predictions are about 2 \GeV.
We are working to complete d+Au reference calculations for the asymmetric case.
Our initial findings indicate energy loss values that are at the lower end of predictions.

%% end of main text

%\section*{Acknowledgments} % please insert, comment out or delete if not needed
%This is where one places acknowledgments for funding bodies etc., if needed.
%For the large collaborations, this is listed once and for all, together with 
%the author lists etc. in the proceedings back-material.

 % do not change 
\end{document}